
\input phyzzx
\PHYSREV
\hoffset=1.25in
\hfuzz=8pt

\def\Re{{\cal R \mskip-4mu \lower.1ex \hbox{\it e}\,}}
\def\Im{{\cal I \mskip-5mu \lower.1ex \hbox{\it m}\,}}
\def\ie{{\it i.e.}}

\def\etal{{\it et al.}}

\def\sub#1{_{\lower.25ex\hbox{$\scriptstyle#1$}}}
\def\sul#1{_{\kern-.1em#1}}
\def\sll#1{_{\kern-.2em#1}}
\def\sbl#1{_{\kern-.1em\lower.25ex\hbox{$\scriptstyle#1$}}}
\def\ssb#1{_{\lower.25ex\hbox{$\scriptscriptstyle#1$}}}
\def\sbb#1{_{\lower.4ex\hbox{$\scriptstyle#1$}}}

\def\tev{\,{\rm TeV}}

\def\to{\rightarrow}
\def\mh{\ifmmode m\sbl H \else $m\sbl H$\fi}
\def\mch{\ifmmode m_{H^\pm} \else $m_{H^\pm}$\fi}
\def\mt{\ifmmode m_t\else $m_t$\fi}
\def\mc{\ifmmode m_c\else $m_c$\fi}
\def\mz{\ifmmode M_Z\else $M_Z$\fi}
\def\mw{\ifmmode M_W\else $M_W$\fi}
\def\mws{\ifmmode M_W^2 \else $M_W^2$\fi}
\def\mhs{\ifmmode m_H^2 \else $m_H^2$\fi}
\def\mts{\ifmmode m_t^2 \else $m_t^2$\fi}
\def\mzs{\ifmmode M^2_Z \else $M^2_Z$\fi}
\def\mcs{\ifmmode m_c^2 \else $m_c^2$\fi}
\def\rlnw{\ifmmode r_{\ell\nu W} \else $r_{\ell\nu W}$\fi}
\def\rnnz{\ifmmode r_{\nu\nu Z} \else $r_{\nu\nu Z}$\fi}
\def\rlnwr{\ifmmode r_{\ell\nu W_R} \else $r_{\ell\nu W_R}$\fi}
\def\mchs{\ifmmode m_{H^\pm}^2 \else $m_{H^\pm}^2$\fi}
\def\zp{\ifmmode Z' \else $Z'$\fi}
\def\ztwo{\ifmmode Z_2\else $Z_2$\fi}
\def\zone{\ifmmode Z_1\else $Z_1$\fi}
\def\mtwo{\ifmmode M_2\else $M_2$\fi}
\def\mone{\ifmmode M_1\else $M_1$\fi}
\def\tb{\ifmmode \tan\beta \else $\tan\beta$\fi}
\def\xw{\ifmmode x\sub w\else $x\sub w$\fi}
\def\ch{\ifmmode H^\pm \else $H^\pm$\fi}
\def\lum{\ifmmode {\cal L}\else ${\cal L}$\fi}
\def\inpb{\ifmmode {\rm pb}^{-1}\else ${\rm pb}^{-1}$\fi}
\def\infb{\ifmmode {\rm fb}^{-1}\else ${\rm fb}^{-1}$\fi}
\def\epem{\ifmmode e^+e^-\else $e^+e^-$\fi}
\def\ppb{\ifmmode \bar pp\else $\bar pp$\fi}

\def\half{\ifmmode {1\over 2} \else ${1\over 2}$\fi}

\newskip\zatskip \zatskip=0pt plus0pt minus0pt
\def\matth{\mathsurround=0pt}
\def\lsim{\mathrel{\mathpalette\atversim<}}

\def\atversim#1#2{\lower0.7ex\vbox{\baselineskip\zatskip\lineskip\zatskip
  \lineskiplimit 0pt\ialign{$\matth#1\hfil##\hfil$\crcr#2\crcr\sim\crcr}}}

%

\catcode`@=12
\catcode`@=11
\def\p@bblock{\begingroup\tabskip=\hsize minus\hsize
   \baselineskip=1.5\ht\strutbox\topspace-2\baselineskip
   \halign to \hsize{\strut ##\hfil\tabskip=0pt\crcr
   \the\Pubnum\cr
   \the\date\cr \the\pubtype\cr}\endgroup}
\Pubnum{\bf ANL-HEP-PR-92-33}
\date={April 1992}
\pubtype={}
\titlepage
\hoffset=1.25in
\title{PROBING NEW GAUGE BOSON COUPLINGS VIA THREE-BODY DECAYS
\footnote{*}{Work supported by the U.S. Department of
Energy, Division of High Energy Physics, Contracts W-31-109-ENG-38
and W-7405-Eng-82.} }

\author{J.L.~Hewett$^a$\footnote{\dag}{Research Supported by an SSC
Fellowship from the Texas National Research Laboratory Commission} \ and \
T.G.~Rizzo$^{a,b}$ }
\smallskip
\address{$^a$High Energy Physics Division, Argonne National Laboratory,
Argonne, IL \ 60439}
\smallskip
\address{$^b$Ames Laboratory, Iowa State University, Ames, IA \ 51010}

\vskip.25in
\abstract

We examine the possibility of using rare, 3-body decays of a new neutral
gauge boson, \ztwo, to probe its gauge couplings at hadron colliders.
Specifically, we study
the decays $\ztwo\to W\ell\nu$ and $\ztwo\to Z\nu\bar\nu$ and find that much
knowledge of the \ztwo\ properties can be obtained from these processes.
In particular, these decay modes can yield valuable information on the amount
of $Z_1-Z_2$ mixing, on the generation dependence of the \ztwo\ couplings,
on the properties of the new generator associated with the \ztwo, as well
as being used to distinguish between possible extended models.  The analogous
3-body decays into a new, heavy charged gauge boson, $\ztwo\to
W_2^\pm\ell^\mp\nu$, are also investigated in models where this can occur.

\endpage
\def\MPL #1 #2 #3 {Mod.\ Phys.\ Lett.\ {\bf#1},\ #2 (#3)}
\def\NPB #1 #2 #3 {Nucl.\ Phys.\ {\bf#1},\ #2 (#3)}
\def\PLB #1 #2 #3 {Phys.\ Lett.\ {\bf#1},\ #2 (#3)}
\def\PR #1 #2 #3 {Phys.\ Rep.\ {\bf#1},\ #2 (#3)}
\def\PRD #1 #2 #3 {Phys.\ Rev.\ {\bf#1},\ #2 (#3)}
\def\PRL #1 #2 #3 {Phys.\ Rev.\ Lett.\ {\bf#1},\ #2 (#3)}
\def\RMP #1 #2 #3 {Rev.\ Mod.\ Phys.\ {\bf#1},\ #2 (#3)}
\def\ZPC #1 #2 #3 {Z.\ Phys.\ {\bf#1},\ #2 (#3)}
\def\IJMP #1 #2 #3 {Int.\ J.\ Mod.\ Phys.\ {\bf#1},\ #2 (#3)}
\Ref\ssc{See, for example, J.L.~Hewett and T.G.~Rizzo, in {\it Proceedings of
the 1988 Snowmass Summer Study on High Energy Physics in the 1990's}, Snowmass,
CO 1988, ed.~S.~Jensen;
V.~Barger \etal, \PRD D35 166 1987 ; L.S.~Durkin and P.~Langacker,
\PLB B166 436 1986 ; F.~Del Aguila, M.~Quiros, and F.~Zwirner, \NPB B287
419 1987 , {\bf B284}, 530 (1987) ;
P.~Chiappetta \etal, to appear in the {\it Proceedings of the Large
Hadron Collider Workshop}, Achen, Germany, 1990.}
\Ref\sdc{Letter of Intent, SDC Collaboration (E.~Berger, \etal),
SDC Report SDC-90-00151 (1990);
I.~Hinchcliffe, M.~Mangano, and M.~Shapiro, SDC Report SDC-90-00036 (1990);
I.~Hinchcliffe, SDC Report SDC-90-00100 (1990); I.~Hinchcliffe, M.~Shapiro,
and J.L.~Siegrist, SDC Report SDC-90-00115 (1990);
G.~Eppley and H.E.~Miettinen, SDC Reports SDC-90-00125 (1990),
and SDC-91-00009 (1991), GEM Letter of Intent (R.~Steiner \etal), (1991).}
\Ref\us{J.L.~Hewett and T.G.~Rizzo, \PRD D45 161 1992 , see also,
K.~Whisnant, to appear in the {\it Proceedings of the 1990 Summer
Study on High Energy Physics - Research Directions for the Decade}, Snowmass,
CO, 1990, ed.~E.L.~Berger; F.~del Aguila and J.~Vidal, \IJMP A4 4097 1989 ;
B.~Aveda
\etal, in {\it Proceedings of the 1986 Summer Study on the Physics of the
Superconducting Supercollider}, Snowmass, CO, 1986 eds.~R.~Donaldson and
J.N.~Marx.}
\Ref\tgr{T.G.~Rizzo, \PLB B192 125 1987 .}
\Ref\penn{M.~Cveti\v c and P.~Langacker, \PRD D46 R14 1992 .}
\Ref\mww{W.~Marciano and D.~Wyler, \ZPC C3 181 1979 .}
\Ref\pb{H.~Plothow-Besch, mini-Rapporteur talk given at the {\it Joint
International Lepton-Photon Symposium and Europhysics Conference on High
Energy Physics}, Geneva, Switzerland, 26 July - 2 August, 1991.}
\Ref\esix{J.L.~Hewett and T.G.~Rizzo, Phys.~Rep.~{\bf 183}, 193 (1989), and
references therein.}
\Ref\lrm{For a review and original references, see, R.N.~Mohapatra, {\it
Unification and Supersymmetry}, (Springer, New York, 1986).}
\Ref\mirjam{For a related discussion, see, M.\ Cveti\v c, P.\ Langacker,
and B.\ Kayser, \PRL 68 2871 1992 ; M.\ Cveti\v c and P.\ Langacker,
\PRD D42 1797 1990 .}
\Ref\kap{D.~Chang, R.~Mohapatra, and M.~Parida, \PRD D30 1052 1984 .}
\Ref\alrm{E.~Ma, \PRD D36 274 1987 ; \MPL A3 319 1988 ; K.S.~Babu \etal,
\PRD D36 878 1987 ; V.~Barger and K.~Whisnant, \IJMP A3 879 1988 ;
J.F.~Gunion \etal, \IJMP A2 118 1987 ; T.G.~Rizzo, \PLB B206 133 1988 .}
\Ref\fh{R.~Foot and O.~Hern\`andez, \PRD D41 2283 1990 ; R.~Foot,
O.~Hern\`andez, and T.G.~Rizzo, \PLB B246 183 1990 , and {\bf B261}, 153
(1991).}
\Ref\blank{K.T.~Mahanthappa and P.K.~Mohapatra, \PRD D42 1732 1990 , and
{\bf D42}, 2400 (1990).}
\Ref\harv{H.~Georgi, E.E.~Jenkins, and E.H.~Simmons, \PRL 62 2789 1989 , and
\NPB B331 541 1990 ;
V.~Barger and T.G.~Rizzo, \PRD D41 946 1990 ; T.G.~Rizzo,
Int.~J.~Mod.~Phys.~{\bf A7}, 91 (1992).}
\Ref\comp{See, for example, R.~Casalbuoni \etal, \PLB B155 95 1985 , and
\NPB B310 181 1988 ; U.~Baur \etal, \PRD D35 297 1987 ; M.~Kuroda \etal,
\NPB B261 432 1985 .}
\Ref\ssm{Such models can occur in some superstring compactification
scenarios, S.~Samuel, private communication.}
\Ref\kuo{A.~Bagneid, T.K.~Kuo, and N.~Nakagawa, \IJMP A2 1327 1987 ,
and {\bf A2}, 1351 (1987).}
\Ref\lept{X.-G.~He, \etal, \PRD D44 2118 1991 .}
\Ref\mastuff{X.~Li and E.~Ma, \PRL 47 1788 1981 , and Univ. of California,
Riverside Report UCRHEP-T90 (1992); E.~Ma, X.~Li, and S.F.~Tuan, \PRL
60 495 1988 ; E.~Ma and D.~Ng, \PRD D38 304 1988 .}
\Ref\lep{For recent combined analysis of LEP data, see P.~Renton, Oxford
University Report OUNP-91-30 (1991); The LEP Collaborations, CERN Report,
CERN-PPE/91-232 (1991).}
\Ref\zww{N.~Deshpande, J.~Gunion, and F.~Zwirner, in the {\it Proceedings
of the Workshop on Experiments, Detectors, and Experimental Areas for the
Supercollider}, Berkeley, CA, 1987, eds.~R.~Donaldson and M.G.D.~Gilchriese;
N.~Deshpande, J.~Grifols, and A.~Mendez, \PLB B208 141 1988 ;
F.~del Aguila \etal, \PLB B201 375 1988 , and {\bf B221}, 408 (1989).}

\endpage


It is now commonly accepted that if a new neutral gauge boson $(Z')$ exists,
it should be observed by direct production, via $pp\to\zp\to\ell^+\ell^-$, at
both the SSC and LHC supercolliders if its mass is of order a few TeV or
less\refmark\ssc\ (provided it couples to both $q\bar q$\ and
$\ell^+\ell^-$ \ pairs at or near electroweak strength).  Indeed, if a \zp\ is
discovered we will want to learn as much about it as possible,
in particular, the next logical step would be to determine its gauge couplings
and the extended model that bears it origin.  Unlike \epem\ machines, hadron
colliders are limited to only a few measurable quantities with which the new
gauge boson properties can be determined.  In addition to obtaining the \zp\
mass, the planned SSC and LHC detectors\refmark\sdc\ will be able to collect
data on the \zp\ production cross section and subsequent decay into $\ell^+
\ell^-$, the full \zp\ width, and the leptonic forward-backward asymmetry.
Unfortunately, these measurements will not only be statistics limited but
also will experience reasonably large systematic effects due to finite mass
resolution and efficiencies as well as uncertainties in
the collider luminosity.  To further extract coupling information,
uncertainties
in the parton distributions will also contribute to the systematic errors.
If, however, several theoretical assumptions are made, one can use
the above data to distinguish new \zp\ bosons from different models with
reasonable reliability\rlap.\refmark\us

In order to obtain more and better information on \zp\ couplings, we need an
additional set of quantities, which do not suffer the large theoretical or
systematic uncertainties discussed above, can be measured with reasonable
statistics, and yet are sensitive to the particular extended model.  Since
decay modes involving leptons provide the cleanest signatures and the
conventional $\ell^+\ell^-$ mode is already being used to discover the $Z'$,
one of the next possibilities to consider are
various three-body decays.  One potential process\rlap,\refmark\tgr\
which has recently been revived\rlap,\refmark\penn\ is to look for the
decay $Z'\to W^\pm\ell^\mp\nu$, and, in particular, to measure the ratio
$$ r_{\ell\nu W}= {\Gamma(Z'\to W^\pm\ell^\mp\nu)\over\Gamma(Z'\to
\ell^+\ell^-)}   \,,   \eqno\eq $$
which suffers very little from the above mentioned systematic uncertainties.
(We note that in this definition of $r_{\ell\nu W}$ we sum over both $W^\pm$
modes, but there is no sum over $\ell$ which we assume to be either $e$
or $\mu$.)   A second such useful quantity\refmark\penn\ is the
corresponding ratio
$$ r_{\nu\nu Z} = {\Gamma(Z'\to Z\nu_\ell\bar\nu_\ell)\over \Gamma(Z'\to
\ell^+\ell^-)} \,, \eqno\eq $$
wherein a sum over the three generations of $\nu_\ell\bar\nu_\ell$ is assumed.
If one allows for decays of the $Z'$ into two jets plus a $W^\pm$ or $Z$, two
additional quantities can be defined which parallel $r_{\ell\nu W}$ and
$r_{\nu\nu Z}$ above.  We feel, however, that though
$W^\pm$ or $Z +$ jets final states from \zp\ decay might be separable
from Standard Model (SM) backgrounds, these modes will no longer be
as clean as the two discussed above.  Thus we
restrict our attention to $r_{\ell\nu W}$ and $r_{\nu\nu Z}$ below, where
we will find that measurements of these two quantities will reveal much
about the nature of the \zp.

We first examine the process $\zp\to W^\pm\ell^\mp\nu$ and the ratio \rlnw.
In general, as discussed in Ref.~\tgr, this reaction can proceed either by
$W$ emission off of a fermion leg, or via a $\zp W^+W^-$ coupling which exists
only if the \zp\ mixes with the SM $Z$.  The Feynman diagrams responsible
for these contributions are displayed in Fig.~1.  If $Z-Z'$ mixing is
non-vanishing,  then both the $Z$ and \zp\ are not mass eigenstates.  The
physical states will then be
$$\eqalign{
Z_2 &= Z'\cos\phi -Z\sin\phi \,, \crr
Z_1 &= Z'\sin\phi +Z\cos\phi \,, \cr } \eqno\eq $$
with the state \zone\ being the one probed at LEP, and \zp($Z$) must be
replaced by \ztwo(\zone) in the discussion above.  We emphasize that the
$\ztwo W^+W^-$ coupling only occurs via this mixing.  Following Ref.~\tgr\
and Marciano and Wyler\rlap,\refmark\mww\ we can then write the quantity \rlnw\
as
$$
\rlnw={G_F\mws\over 2\sqrt 2\pi^2} (v^2_{2\ell }+a^2_{2\ell } )^{-1}
\left\{\eqalign{
& {1\over 2} [(v_{2\ell}+a_{2\ell})^2+(v_{2\nu}+a_{2\nu})^2]H_1 \crr
&+(v_{2\ell}+a_{2\ell})(v_{2\nu}+a_{2\nu})H_3 \crr
&+{1\over 2} (-s_\phi c^2_w)^2H_2 \crr
&-s_\phi c^2_w
[(v_{2\ell}+a_{2\ell})-(v_{2\nu}+a_{2\nu})]H_4 \cr}\right\}\,,   \eqno\eq $$
where the $v's$ \ and $a's$ \ represent the various vector and axial vector
couplings of the \ztwo\ to charged leptons and neutrinos, $s_\phi
=\sin\phi$, and $c_w=\cos\theta_w$, with $x_w=\sin^2\theta_w$.   Note that
the last two terms in this expression are proportional to the amount of
$Z-Z'$ mixing and arise from the diagram of Fig.~1c.   The quantities
$H_i$ are the results of performing one
dimensional integrations over modified forms of the functions given in
Ref.~\mww.  These modifications arise for $H_{2,4}$ only (the terms that arise
from the $Z_2W^+W^-$ coupling), since we must now integrate over the
$W$-resonance requiring that the finite $W$-width, $\Gamma_W$,
be included in the calculation.  This was not included in the analysis of
Ref.~\mww\ since both of the $W's$ could not be on-shell simultaneously as
$M_Z <2M_W$.  Thus the $H_i$ functions depend only on $M_W$, $\Gamma_W$, and
the \ztwo\ mass, \mtwo.  Clearly, \rlnw\ will be quite sensitive to
$s_\phi\neq 0$; when $s_\phi=0$, only the above terms with $H_{1,3}$ will
remain.  {\it For the moment}, we will assume that $s_\phi=0$, and
will neglect any possible influence from $W-W'$ mixing.  In Fig.~2 we present
the number of events expected per year at the SSC with an integrated luminosity
of $10^4~\inpb$ from the process $pp\to\ztwo\to W^\pm\ell^\mp\nu$ as a function
of the \ztwo\ mass for various extended models, which are discussed below.
We see that hundreds of events are expected for \ztwo\ masses up to $\sim
2\tev$ in most models.  Here we have included a lepton identification
efficiency\refmark\sdc\ of $\epsilon=85\%$ for each lepton.

Since it is assumed that a \ztwo\ exists,
it must couple to a new diagonal generator, $D$, originating from an extended
gauge group.  If $D$ and the ordinary $SU(2)_L$ generator $T_{iL}$ commute,
\ie, $[D, T_{iL}]=0$, or if only
$$ [D,T_{iL}]| ~\nu_L,\ell_L>=0 \eqno\eq $$
is satisfied, then $v_{2\ell}+a_{2\ell}=v_{2\nu}+a_{2\nu}$ and \rlnw\
simplifies to
$$ \rlnw={G_F\mws\over 2\sqrt 2\pi^2} (H_1+H_3)(v_{2\ell}+a_{2\ell})^2
(v^2_{2\ell}+a^2_{2\ell})^{-1} \,.  \eqno(4') $$
Note that Eq. ($4'$) can never be satisfied when $s_\phi\neq 0$ since $D$ will
then contain the term $-s_\phi(T_{3L}-x_wQ)$ and neither $Q$ nor $T_{3L}$
commutes with $T_{1,2L}$.  We will return to Eqs.~(4-5) after a brief
discussion of \rnnz.

Let us also examine the ratio $r\equiv\Gamma(\ztwo\to\zone f\bar
f)/\Gamma(\ztwo
\to\ell^+\ell^-)$, which is given by
$$ r= {G_F\mzs\over 8\sqrt 2\pi^2} N_cN_f {(v_{1f}+a_{1f})^2(v_{2f}+a_{2f})^2
+ (v_{1f}-a_{1f})^2(v_{2f}-a_{2f})^2\over (v^2_{2\ell}+a^2_{2\ell})} ~~I \,,
\eqno\eq $$
where $N_c$ is the usual color factor, $I$ is a two-dimensional parameter
integral which depends only on $M_1^2/M_2^2$ (when fermion masses are
neglected), and $N_f$ labels the number of flavors of a given type.  For
three generations of left-handed neutrinos, $N_c=1$, $N_f=3$, $v_{1\nu}=
a_{1\nu}$, and $v_{2\nu}=a_{2\nu}$ so that
$$ \rnnz= {3G_F\mzs\over 2\sqrt 2\pi^2}{4v^2_{1\nu}v^2_{2\nu}\over
(v^2_{2\ell}+a^2_{2\ell})} ~~I \,. \eqno\eq $$
Here we have assumed that all the various couplings are generation
independent in performing the sum over $\nu_e$, $\nu_\mu$, and $\nu_\tau$.
Note that with our normalization convention, $4v^2_{1\nu}=1$ when $s_\phi=0$.
We anticipate that, unlike \rlnw, \rnnz\ will not be greatly affected by
$s_\phi\neq 0$; we will see further below that this is the case.  For now, we
continue to assume that $s_\phi=0$ and also take Eq.~($4'$) to be valid,
we then see that
$$  \rnnz= K_Z ~~{v^2_{2\nu}\over v^2_{2\ell}+a^2_{2\ell}} \,. \eqno\eq $$
and, using the fact that $v_{2\nu}+a_{2\nu}=v_{2\ell}+a_{2\ell}$
together with $a_{2\nu}=v_{2\nu}$ we find
$$ \rlnw= K_W ~~{v^2_{2\nu}\over v^2_{2\ell}+a^2_{2\ell} } \,, \eqno\eq $$
Here $K_{W,Z}$ are functions of the gauge boson masses only, (\mtwo, \mw,
and \mone) and are {\it independent} of the choice of extended electroweak
model.  Thus, if the above conditions hold, all predictions for $\rnnz/\rlnw$
must lie on a straight line, \ie,
$$ {\rnnz\over \rlnw}={K_Z\over K_W} \,.  \eqno\eq $$
Furthermore, Eqs.~(8) and (9) tell us that {\it both \rnnz\ and \rlnw\ are
bounded}
$$ \eqalign{
 0 &\leq\rlnw\leq\half K_W \,, \crr
 0 &\leq\rnnz\leq\half K_Z \,, \cr  }\eqno\eq $$
with the lower (upper) end-points of these ranges occurring for a purely
right-handed (left-handed) \ztwo\ coupling to leptons.  Thus not only
is the ratio
$\rnnz/\rlnw$ model independent, but the value of the quantities
themselves are restricted to a small region of the $\rnnz - \rlnw$ plane,
with both being dictated {\it solely} by the values of $M_{1,2}$ and \mw.
In addition, the position of the measured values of \rnnz\ and \rlnw\ along
the line will yield information on the ratio of the vector and axial-vector
couplings of the \ztwo, up to a two-fold ambiguity, $v_\ell\leftrightarrow
a_\ell$.

As an application of these results, we now examine the $\rnnz - \rlnw$ plane
for some of the more well-known extended gauge models, taking $\mtwo=1$ TeV for
purposes of demonstration.  We also use $\mone=91.175$ GeV, $M_W=80.14$
GeV, $\Gamma_W=2.15$ GeV (Ref.~\pb), and $x_w=0.2330$ in our numerical analysis
below.  We stress that all these results assume $s_\phi=0$.

Most extended models have generation independent couplings and have generators
satisfying Eq.~(5), thus predicting that the set of values for \rnnz\ versus
\rlnw\ lie on a bounded line segment; this is seen explicitly in Fig.~3.  The
solid line indicates the range of values permitted in the superstring-inspired
effective rank-5 model (ER5M)\refmark\esix, where the \ztwo\ couplings depend
upon a parameter $-90^\circ\leq\theta\leq 90^\circ$.  In the figure, $\psi$
labels the point $\theta=0^\circ$, whereas $\chi$ labels the point $\theta=\pm
90^\circ$ in these models.  In the Left-Right Symmetric Model
(LRM)\refmark\lrm\
the only free parameter is the ratio of the $SU(2)_{L,R}$ couplings,
$\kappa=g_R/g_L$\rlap.\refmark\mirjam\
Note that on general grounds, it is expected\refmark\kap\
that $\kappa\leq 1$.  L labels the point in the figure where $g_R/g_L=1$,
while the extreme case of $\kappa^2=x_w(1-x_w)^{-1}$ coincides with the point
$\chi$.  The position of the prediction for the Alternative Left-Right Model
(ALRM)\refmark\alrm\ is labeled by A, and that of the Foot-Hernandez (FH)
Model\refmark\fh\ is labeled by F.  The expectations for the \ztwo\ model of
Mahanthappa and Mohapatra\refmark\blank, where the new generator $D$ is
proportional to $Y/2$, coincides with those of FH.  Although all these models
are quite different, their predictions for the ratio $\rnnz/\rlnw$ {\it are
found to lie on a straight line within the bounded region} as expected.

Other models shown in Fig~3 demonstrate how Eqs.~(10) and (11) can be violated
if certain conditions are met.  Several models predict that the new generator,
$D$, will not satisfy Eq.~(5), particularly if the \ztwo\ couplings are
proportional to $T_{3L}$.
In the Un-unified model of Georgi \etal\refmark\harv\ (labeled by H in the
figure), $D\sim t_\chi T^\ell_{3L}-T^q_{3L}t_\chi^{-1}$ where $t_\chi=\tan\chi$
with $\chi$ being a mixing parameter, and $T_{3L}^{\ell,q}$ are the third
components of the lepton and quark isospin generators.  Clearly,
Eq.~(4) and thus Eqs.~(10) and (11) are not satisfied in this case. This can
also happen in some compositeness based \ztwo\ models\refmark\comp\ or ones
which predict that the \ztwo\ is just a heavier version\refmark\ssm\ of the
\zone; in the latter case, the prediction is marked by `S' in the figure, while
a \ztwo\ whose couplings directly depend on $T^\ell_{3L}$
will occupy the same position on the figure as in the model of Georgi \etal.

A second possible source of deviation from the straight line prediction of
Eq.~(10) arises from the additional assumption used in Eq.~(7) that the
leptonic couplings of the \ztwo\ are generation independent.  In the model of
Kuo and collaborators\refmark\kuo, the third generation couples differently
than
the first two, whereas, in the Leptophilic model\refmark\lept, where
differences
in lepton number are gauged, the third generation decouples completely from the
\ztwo.  The expected values of \rlnw\ and \rnnz\ in these two models are
labeled by K and E, respectively, in Fig.~3.  In the Leptophilic case, the
values shown in the figure are only for purposes of demonstration, since
this \ztwo\ cannot be produced at a hadron collider.  As a last example, the
predictions from the model of Li and Ma\rlap,\refmark\mastuff\ which also
results in a violation of universality, are found to lie along the vertical
dashed curve with the particular position being dependent upon the value of a
model parameter, $p$.  For $p=1/3$, the model of Kuo \etal\
is recovered.  In fact, one finds that the expectations in all models with
generation dependent couplings and with $D\sim T^\ell_{3L}$ lie along
this dashed line, labeled by `M' in Fig.~3.  None of these
models will generate values of \rnnz\ and \rlnw\ which lie on the straight
line predicted in Eq.~(10).

To summarize our results so far, we have observed that if the following
conditions hold:
$$\eqalign{
 (i)   &~~s_\phi=0 \,, \crr
 (ii)  &~~[D, T_{iL}]|~\nu_L,\ell_L>=0 \,, \crr
 (iii) &~~v_{2f}~~{\rm and}~~a_{2f}~~{\rm are~generation~independent} \,, \cr}
\eqno\eq $$
{\it then and only then} will $\rnnz/\rlnw$ be model independent and both
ratios
be separately bounded by $\half K_{W,Z}$.  Thus, if a \ztwo\ is discovered and
its corresponding values of \rnnz\ and \rlnw\ are determined, and it is
observed that these values lie `elsewhere' on the $\rnnz - \rlnw$ plane rather
than along the solid line, one can safely conclude that at least
one of the above conditions (i)-(iii) are not valid.  We have seen, however,
that for $s_\phi=0$  only rather `exotic' extended models, which do not arise
from conventional grand unified theories, fail to satisfy these conditions.

As a final point of this discussion, we stress that a measurement of \rlnw\
and \rnnz\ alone can not uniquely determine the model of origin of the \ztwo.
This can be seen clearly from Fig.~3, {\it e.g.}, in the case where the LRM
and a particular value of $\theta$ from the ER5M predict the same pair of
values for \rlnw\ and \rnnz.  Even
within the ER5M itself, except for the cases where $\rnnz=\rlnw=0$ and
$\rnnz=\half K_Z$, $\rlnw=\half K_W$ (\ie, the two endpoints of the line), each
point along the line corresponds to two distinct values of the $\theta$
parameter resulting from the $v_\ell\leftrightarrow a_\ell$ ambiguity
mentioned above.
Thus, other data will be required to uniquely determine the origin of the
\ztwo.
We note that the leptonic forward-backward asymmetry (in the narrow width
approximation) at hadron colliders is also invariant when the vector and
axial-vector couplings of the \ztwo\ are flipped for both quarks and leptons.
We also mention in passing that, as discussed in Ref.~\penn, not much
information can be gained by considering the ratio $r_{\ell\ell Z}$
($\equiv r$ in Eq.~(6) with $f=\ell$).  In this case we find
$$\eqalign{
 r_{\ell\ell Z} &= K'_Z \Bigl[ 1+\bigl(
{ 2v_{1\ell }a_{1\ell }\over  v^2_{1\ell }+a^2_{1\ell } } \bigr)
\bigl( { 2v_{2\ell}a_{2\ell}\over v^2_{2\ell}+a^2_{2\ell} }\bigr) \Bigr] \crr
&= K'_Z \Bigl[ 1+0.135\bigl(
{ 2v_{2\ell}a_{2\ell}\over v^2_{2\ell}+a^2_{2\ell} }
\bigr)\Bigr] \,, \cr }\eqno\eq $$
where $K'_Z$ is again a model independent constant and the last equality holds
for $s_\phi=0$ and $x_w=0.2330$.  The sensitivity to coupling variations in
$r_{\ell\ell Z}$ is thus seen to be substantially reduced compared to both
\rlnw\ and \rnnz.

Next we examine what happens when a \ztwo, which satisfies  conditions (ii)
and (iii) above when $s_\phi=0$, is now allowed to mix with the SM $Z$, \ie,
what happens when $s_\phi$ is non-zero.  Clearly, condition (iii) remains
valid,
but if (i) is violated so is (ii), as the new generator $D$ now has a term
proportional to $s_\phi T_{3L}$.  For the case of \rlnw, both terms $H_2$ and
$H_4$ in Eq.~(4) will now contribute. To be specific, we examine the effect of
$s_\phi\neq 0$ in the ER5M, ALRM, and LRM (with $\kappa=1$); all of which
satisfy conditions (ii) and (iii) when $s_\phi=0$.  We first summarize some
properties of the $Z-Z'$ mixing mechanism before discussing its effect on
\rlnw\ and \rnnz.

For an extended model with Higgs scalars transforming only as $SU(2)_L$
doublets or singlets, the $Z-Z'$ mass matrix can be written as
$$\left(\matrix{M^2_Z&\gamma M^2_Z\cr
          \gamma M^2_Z&M^2_{Z'}\cr}\right) \,, \eqno\eq $$
with $\gamma$ being a model dependent parameter of order unity and \mz\ the
value of the SM Z-boson mass in the absence of mixing.  The eigenvalues of
this matrix, $M^2_{1,2}$, correspond to the masses of the physical gauge
bosons, $Z_{1,2}$, given in Eq.~(3).    Since \mone\ is known
from LEP\refmark\lep\ ($=91.175$ GeV), the value of $\phi$ is calculable from
the above Eq.~(14), for a given value of the \ztwo\ mass, \mtwo, in a
particular
model (which then determines $\gamma$).  We can write
$$\eqalign{
M^2_Z &= {M^2_1+M^2_2-[(M^2_2-M_1^2)^2-4\gamma^2M_1^2M_2^2]^{1/2}\over
2(1+\gamma^2)} \,, \crr
M^2_{Z'} &= M_1^2+M_2^2-M^2_Z \,, \cr }\eqno\eq $$
so that one obtains
$$ \phi(M_2,\gamma)=\half\tan^{-1}\Biggl( {2\gamma M^2_Z\over M^2_Z-M^2_{Z'}}
\Biggr) \,. \eqno\eq $$
For the various models we consider, $\gamma$ is given by
$$\eqalign{
\gamma_{LRM} &= -(1-2x_w)^{1/2} \,, \crr
\gamma_{ALRM} &= {x_wt^2_\beta-(1-2x_w)\over (1-2x_w)^{1/2}(1+t^2_\beta)}
\,, \crr
\gamma_{ER5M} &= -2\sqrt{{5x_w\over 3}}\Bigl[\Bigl( {c_\theta\over\sqrt 6}
-{s_\theta\over\sqrt 10}\Bigr) t^2_\beta - \Bigl( {c_\theta\over\sqrt 6}
+{s_\theta\over\sqrt 10}\Bigr)\Bigr] (1+t^2_\beta)^{-1} \,, \cr }\eqno\eq $$
where $t_\beta=\tan\beta=v_t/v_b$, the usual ratio of vacuum expectation
values (vev's) responsible for the top and bottom quark masses, and $s_\theta(
c_\theta)=\sin\theta(\cos\theta)$ being the ER5M mixing angle discussed above.
Note that if $\theta=-90^\circ$ (model $\chi$) then $\gamma_\chi=
-(2x_w/3)^{1/2}$ is independent of the value of $\tan\beta$.  In obtaining
these expressions, we have
made the following assumptions: for the ER5M and ALRM which are based on
superstring-inspired $E_6$, we assume that the only scalars responsible for
$SU(2)_L$ breaking are the SUSY partners of the exotic fermions $N$ and $N^c$
that lie in the {\bf 27} representation.  Since the quantum numbers of
these fields are fixed (for a given value of $\theta$ in the ER5M case), this
completely determines $\gamma$ except for the vev ratio, $\tan\beta$.  In the
LRM case, assuming that the left-handed triplet vev is small implies that the
fields in the `mixed-doublet' representation, $(1/2,1/2)$ of $SU(2)_L\times
SU(2)_R$ are mainly responsible for $SU(2)_L$ breaking.  In this case, the
$\tan\beta$ dependence factors out and one is left with the above expression.
We have for the moment also ignored the possible influence of $W-W'$ mixing in
the LRM case.

Since $\gamma$ is independent of $\tan\beta$ in both model $\chi$ and the LRM,
the value of $s_\phi$ is then uniquely determined in these cases once \mtwo\
is specified.  Figures 4a-b display \rlnw\ as a function of \mtwo\ for (a)
model $\chi$ and (b) the LRM with $\gamma$ as given above (\ie, $s_\phi\neq 0$)
and, for comparison, with $s_\phi=0$ set by hand. Clearly the effects of
$s_\phi\neq 0$ on \rlnw\ are quite striking as it produces a very substantial
increase in the value of this parameter.  (This result was anticipated quite
some time ago in Ref.~\tgr.)  Since $\gamma$ is $\tan\beta$ dependent in the
other models, we present \rlnw\ as a function of $\tan\beta$ in Fig.~5a,
assuming $\mtwo=1$ TeV, for the
ALRM and the three ER5M's corresponding to $\theta=0^\circ$ (model $\psi$),
$\theta=\sin^{-1}\sqrt{3/8}\simeq 37.76^\circ$ (model $\eta$), and $\theta=
-\sin^{-1}\sqrt{5/8}\simeq -52.24^\circ$ (model I).  In all cases, varying
$\tan\beta$ away from the point where $\gamma=0$, \ie, $\tan\beta=1(2,\simeq
1.5)$ for model $\psi$ (model $\eta$, ALRM), which corresponds to $s_\phi=0$,
can produce a substantial increase in the value of \rlnw.  A minimal value of
\rlnw, corresponding to a choice of $\tan\beta$ which produces $s_\phi=0$, will
exist for all values of $\theta$ in the range $-\sqrt{5/3}\leq\tan\theta\leq
\sqrt{5/3}$.  The value of $\tan\beta$ which yields these minimas is given by
$$ \tan^2\beta={1+\sqrt{3/5}\tan\theta\over 1-\sqrt{3/5}\tan\theta} \,.\eqno\eq
$$
Thus, for example, model I, with $\tan\theta=-\sqrt{5/3}$, will not experience
any true minima of \rlnw\ for finite values of $\tan\beta$.
This is demonstrated in Fig~5b, which is a three-dimensional plot of \rlnw\
as a function of $\tan\beta$ and $\theta$, where a minima "valley" is clearly
observable.  We conclude that even though
$|\phi |\lsim 10^{-3}$ for \mtwo\ in the TeV range, this small amount
of mixing can substantially modify the expectations for the value of \rlnw\
within a given model.

How does $Z-Z'$ mixing modify the values of \rnnz?  We anticipate that there is
little effect since $s_\phi\neq 0$ does not induce a resonant contribution to
this ratio.  Hence, for this case, the inclusion of mixing only results in a
slight shift of the gauge boson coupling constants.  Figures 6a-b show \rnnz\
as a function of \mtwo\ for (a) model $\chi$ and (b) the LRM, and demonstrate
that our expectations are correct.  Thus, for models which satisfy conditions
(ii) and (iii) of Eq.~(12) when $s_\phi=0$, the cleanest signal for
$s_\phi\neq 0$ is that \rlnw\ would be substantially increased while \rnnz\
would suffer only a slight modification.  This would correspond
to a shift of the model predictions to the right and off of the straight line
in Figure 1.  If \rnnz\ and \rlnw\ were the only properties of the
\ztwo\ that were measured, this would imply that it would be impossible to
separate a model which violates conditions (ii) and (iii) with $s_\phi=0$
from a model which is shifted off of the straight line due to a non-zero
value of $s_\phi$.  As an example, the Leptophilic model \ztwo\ would be
indistinguishable from an ER5M \ztwo\ with $\theta\simeq 10^\circ$
and with a value of $\tan\beta$ which increases \rlnw\ (via $s_\phi
\neq 0$) by a small amount.  However, the observation of a violation of the
bound on \rnnz\ in Eq.~(11), would clearly signal that the conditions (ii) or
(iii) are violated {\it independently} of whether $s_\phi=0$ or not.  Thus,
while \rlnw\ is the more sensitive probe for the validity of condition (i), the
ratio \rnnz\ does the corresponding job of testing the validity of conditions
(ii) and (iii).  Combining knowledge of the values of \rlnw\ and \rnnz\ with
the
measured values of the relative branching fractions for the processes
$\ztwo\to\epem, \mu^+\mu^-$, and $\tau^+\tau^-$ would completely determine
the validity of any of these conditions.

Up to this point, we have ignored the possibility that a new charged gauge
boson, $W_2^\pm$, may also participate in 3-body \ztwo\ decays.  New charged
gauge bosons are present in several of
the models discussed above, in particular, the LRM\rlap,\refmark\lrm\
ALRM\rlap,\refmark\alrm\ Li and Ma model\rlap,\refmark\mastuff\ and
HARV model\rlap.\refmark\harv\  In the Li and Ma and HARV cases,  the \ztwo\
and $W_2^\pm$ are essentially degenerate so that $W_2^\pm$ final states in
\ztwo\ decay are uninteresting.  This is not generally the case for either the
LRM or ALRM, where $\ztwo\to W_2^\pm\ell^\mp\nu_R$ is always kinematically
accessible. The two body decay $\ztwo\to W_2^+W_2^-$ might also be allowed in
the LRM for a certain range of the model parameters.  To be concrete, we will
neglect any effects associated with $W-W'$ mixing (which is naturally absent
in the ALRM) and $Z-Z'$ mixing.  The \ztwo\ to $W_2^\pm$ mass ratio is
$$ {M^2_{Z_2}\over M^2_{W_2}} =
{\kappa^2(1-x_w)\over \kappa^2(1-x_w)-x_w} ~~\rho_R  \,, \eqno\eq $$
where $\kappa\equiv g_R/g_L$ is the ratio of $SU(2)_{L,R}$ couplings, and
$\rho_R$ probes the symmetry breaking sector relevant for the heavy
gauge boson pair:
$$\rho_R\equiv {{2{\sum_i} T^2_{3R_i}{\rm v^2_i}}\over {{\sum_i}[T_{R_i}
(T_{R_i}+1)
-T^2_{3R_i}]{\rm v^2_i}}} = \cases{1,&{\rm Higgs~doublets};\cr
                                   2,&{\rm Higgs~triplets}.\cr} \eqno\eq $$
Here, the sum extends over the Higgs sector, ${\rm v_i}$ is the vev
of the $i^{th}$ Higgs boson, and $T_{R_i}(T_{3R_i})$ is the value of isospin
(third component of isospin) of the neutral Higgs boson under
$SU(2)_R$\rlap.\refmark\mirjam\
In the LRM, \ $0.55\lsim\kappa\lsim 1$\ and \ $\rho_R$\ takes either value
depending on whether the neutrinos are Majorana or Dirac particles, whereas
in the ALRM, $\kappa=\rho_R=1$ only.  The two-body decay $\ztwo\to W_2^+W_2^-$
is kinematically accessible in the LRM for the range $\kappa\lsim 0.63(0.77)$
with a doublet (triplet) $SU(2)_R$ symmetry breaking sector.  Figure 7 displays
the ratio $M_{Z_2}/M_{W_2}$ as a function of $\kappa$ for both Higgs
doublet and triplet representations.

Denoting the $\ztwo W_2^+W_2^-$ coupling as $\lambda g_R$, we can define a
ratio similar to \rlnw\ above,
$$\eqalign{
r_{\ell\nu W_R} &\equiv {\Gamma(\ztwo\to W_2^\pm\ell\nu_R)\over
\Gamma(\ztwo\to\ell^+\ell^-) } \crr
&= {G_F\mws\over 2\sqrt 2\pi} (v^2_{2\ell} + a^2_{2\ell})^{-1}
\Bigg\{{1\over 2} [(v_{2\ell}-a_{2\ell})^2+(v_{2\nu_R}-a_{2\nu_R})^2]H'_1 \crr
&\qquad\qquad\qquad\qquad\qquad\qquad
+(v_{2\ell}-a_{2\ell})(v_{2\nu_R}-a_{2\nu_R})H'_3\crr
&\qquad\qquad\qquad\qquad\qquad\qquad  + \half \kappa^2\lambda^2H'_2\crr
&\qquad\qquad\qquad\qquad\qquad\qquad
 +\kappa\lambda [(v_{2\ell}-a_{2\ell})-(v_{2\nu_R}-a_{2\nu_R})]H'_4
\Bigg\} \,, \cr } \eqno\eq $$
where $H'_i=H_i$ with the replacements $\mw\to M_{W_2}$, $\Gamma_W\to
\Gamma_{W_2}$.  All four terms will contribute to \rlnwr\ since the couplings
are always linearly proportional to $T_{3R}$ (\ie, the condition
$v_{2\ell}-a_{2\ell}=v_{2\nu_R}-a_{2\nu_R}$ doesn't hold in the LRM or
the ALRM).  The parameter $\lambda$ introduced above is given by
$$ \lambda = \Bigg[ {\kappa^2-(1+\kappa^2)x_w\over \kappa^2(1-x_w)}\Bigg]^{1/2}
\,, \eqno\eq $$
which is simply $M_{W_2}/M_{Z_2}$ for $\rho_R=1$ (in analogy with the factor
$c_w=M_W/M_Z$ which is present in the SM trilinear coupling).
We note that the expression for
\rlnwr\ assumes that $\nu_R$ is light relative to the \ztwo\ and $W_2^\pm$;
this is an excellent approximation in the LRM with a doublet Higgs
representation and in the ALRM where $"\nu_R"$ is expected to be light (the
exotic fermion $S^c_L$ in the {\bf 27} representation of $E_6$, Ref.~\esix,
plays the role of the right-handed neutrino in the ALRM).  For completeness,
we note that the rate for \ztwo\ decay to an
{\it on-shell} pair of $W_2$'s is given by
$$\eqalign{
\Gamma(\ztwo\to W_2^+W_2^-) &= {G_F\mws\over 24\sqrt 2\pi} M_{Z_2}
\lambda^2\kappa^2 \Bigg( {M^2_{Z_2}\over M^2_{W_2}}\Bigg)^2
\Bigg( 1-{4M^2_{W_2}\over M^2_{Z_2}}\Bigg)^{3/2}\crr
&\quad\times \left\{1+20\Bigg( {M^2_{W_2}\over M^2_{Z_2}}\Bigg) +12\Bigg(
{M^2_{W_2}\over M^2_{Z_2}}\Bigg)^2 \right\}   \,. \cr} \eqno\eq $$
This width is potentially quite large for smaller values of $\kappa$, as in
this case $M^2_{Z_2}\gg M^2_{W_2}$ and no mixing angle suppression appears.
Figure 8 presents the reduced width $\Gamma_R\equiv\Gamma(\ztwo\to W_2^+W_2^-)
/M_{Z_2}$ as a function of $\kappa$ for both types of symmetry breaking
sectors; note that in order to set the scale and guide the eye, the
corresponding ratio for the SM $Z$ decay into
$e^+e^-$ is $\simeq 9.1\times 10^{-4}$.  We see from the figure that
$\Gamma_R$ is only significant for the doublet Higgs representation when
$\kappa\lsim 0.61$, but remains much larger in the triplet case out to
values of $\kappa\simeq 0.73$.

Figure 9 shows the ratio \rlnwr\ as a function of $\kappa$ for both the
triplet and doublet symmetry breaking schemes assuming $M_{Z_2}=4\tev$ for
purposes of demonstration.   (This choice of $M_{Z_2}$ was made in order to
avoid too light a value of $M_{W_2}$ for small $\kappa$.)  In the Higgs
triplet case, \rlnwr\ remains above $10^{-2}$ for almost all the entire range
of $\kappa$, whereas, the ratio drops below this value for $\kappa\simeq 0.63$
in the case of scalar doublets.  The very large value of \rlnwr\ at small
$\kappa$ values arises from the strong resonant contribution, $\ztwo\to
W_2^+W_2^-$, in a manner similar to what we saw above in the case of $Z-Z'$
mixing for $\ztwo\to W_1^+W_1^-$.  Similar results are obtainable for other
values of $M_{Z_2}$.  For the ALRM, where $\kappa=\rho_R=1$,
\rlnwr\ is found to be extremely small and unobservable, \ie, $\lsim 10^{-4}$.

As mentioned above, one could also gain information\refmark\penn\
from the decays,
$\ztwo\to W+$ jets and $\ztwo\to Z+$ jets, however these particular processes
will suffer from more severe SM backgrounds, such as $W$ or $Z + n-jet$
production.  Not only are the leptonic processes, $\ztwo\to W\ell\nu$ and
$\ztwo\to Z\nu\bar\nu$, cleaner to begin with, but their kinematic
distributions should be able to differentiate them from SM backgrounds such
as $pp\to ZZ, WW$, as well.  The fermions in the decay $\ztwo\to f\bar f$
will come out relatively back to back and the gauge boson, which is
bremsstrahlunged off of one of the fermion legs, will be approximately
collinear with the
fermion and relatively soft.  The resonance graph, $\ztwo\to W^+W^-$ will
have different kinematical properties\rlap.\refmark\zww\
Detailed background studies are clearly needed, but are beyond the scope
of this work.

In summary, we have examined the 3-body decays, $\ztwo\to W\ell\nu$ and
$\ztwo\to Z\nu\bar\nu$ and have found that they can be used to obtain much
information on the properties of the \ztwo\ for $M_2\lsim 2-3\tev$.
Besides being used to differentiate between possible extended gauge models,
these processes can measure the amount of $Z-Z'$ mixing, the generation
dependence of the \ztwo\ couplings, and the properties of the new generator
associated with the \ztwo.  In particular, if the \ztwo\ arises from a
more "conventional" grand unified theory and $Z-Z'$ mixing is absent, the
predictions for the 3-body decays lie on a straight line in the \rlnw\ - \rnnz\
plane, with the slope of the line being determined by the mass of the \ztwo.
If any of the conditions stated in Eq.~(12) are violated, then the values
of these decay rates will not lie on this line.  The effect of $Z-Z'$
mixing is to increase the rate for \rlnw, while keeping the prediction
for \rnnz\ relatively unchanged.  Hence, a measurement of $\rnnz>\half K_Z$
is a definite signal for the violation of conditions (ii) and (iii) of
Eq.~(12), while a measurement of $\rlnw>\half K_W$ could also be a signature
for non-zero $Z-Z'$ mixing.  We also find that the decays into a new
heavy charged gauge boson, $\ztwo\to W_2^\pm\ell^\mp\nu$, can occur in
some models at observable rates and would yield even more information on
the origin of the extended gauge sector.

We urge our experimental colleagues to consider these promising 3-body
decays!

\vskip.25in
\centerline{ACKNOWLEDGEMENTS}

The research of J.L.H. was supported in part by awards granted by the Texas
National Research Laboratory Commission and by the U.S.~Department of Energy
under contract W-31-109-ENG-38.  The research of T.G.R. was supported by
the U.S.~Department of energy under contracts W-31-109-ENG-38
and W-7405-ENG-82.

\endpage

\refout
\endpage

%
\FIG\one{Feynman diagrams responsible for the decay $Z_2\to\ell\nu W$.}
\FIG\two{Number of events expected for the process $Z_2\to \ell\nu W$
neglecting
$Z-Z'$ mixing at the SSC with $10^4~\inpb$ of integrated luminosity as a
function of the \ztwo\ mass.
{}From top to bottom, the dashed-dotted curve corresponds to the SSM, the
dashed curve to the HARV model (with $s_\phi=0.5$), the dotted curve to the
ALRM, the solid curve to the ER5M $\chi$, and the short-dotted curve to the
LRM.}
\FIG\three{Values of \rnnz\ and \rlnw\ predicted by the various models
discussed in the text when $s_\phi=0$.}
\FIG\four{A comparison of the predicted values of the ratio \rlnw\ in (a)
model $\chi$ and (b) the LRM as a function of \mtwo\ both with (solid
curve) and without (dashed curve) $Z-Z'$ mixing.}
\FIG\five{(a)  \rlnw\ as a function of $\tan\beta$ assuming $\mtwo=1\tev$ for
the $E_6$ ER5M I (corresponding to $\theta=-52.24^\circ$), represented by the
solid curve; model $\eta$ ($\theta=37.76^\circ$), dotted curve, and model
$\psi$ ($\theta=0^\circ$), dashed curve;  as well as the ALRM, dash-dotted
curve.  (b)  Three-dimensional figure of \rlnw\ as a function of $\tan\beta$
and $\theta$ in the ER5M.  The x-axis corresponds to $\theta$ (ranging from
$-100^\circ$ to $+100^\circ$), the y-axis to $\tan\beta$ (ranging from
$10^{-1}$
to $10^1$), and the z-axis to \rlnw\ (ranging from 0 to 5).}
\FIG\six{Same as Fig.~4 but for the ratio \rnnz.}
\FIG\seven{The ratio $M_{Z_2}/M_{W_2}$ as a function of $\kappa$ for a
triplet (solid curve) or doublet (dashed-dot curve) symmetry breaking sector.}
\FIG\eight{The reduced width, $\Gamma_R$, for the same cases displayed in
Fig.~7.}
\FIG\nine{The ratio \rlnwr\ for the same Higgs representations as
shown in Fig.~7.}

\figout\endpage

\bye